# Green-Kubo formulas with symmetrized correlation functions for quantum systems in steady states: the shear viscosity of a fluid in a steady shear flow

Hiroshi Matsuoka

*Department of Physics, Illinois State University, Normal, Illinois, 61790-4560, USA*

Phone: (309) 438-3236

FAX: 309-438-5413

hmb@phy.ilstu.edu

## ABSTRACT

For a quantum system in a steady state with a constant current of heat or particles driven by a temperature or chemical potential difference between two reservoirs attached to the system, the fluctuation theorem for the current was previously shown to lead to the Green-Kubo formula for the linear response coefficient for the current expressed in terms of the *symmetrized* correlation function of the current density operator. In this article, we show that for a quantum system in a steady state with a constant rate of work done on the system, the fluctuation theorem for a quantity induced in the system also leads to the Green-Kubo formula expressed in terms of the *symmetrized* correlation function of the induced quantity. As an example, we consider a fluid in a steady shear flow driven by a constant velocity of a solid plate moving above the fluid.

# 1. INTRODUCTION AND A SUMMARY OF OUR RESULTS

In the standard linear response theory [1] for a quantum system subject to a time-dependent external field, the Green-Kubo formula for the linear response coefficient for a quantity induced by the external field is expressed in terms of the *canonical* correlation function involving the induced quantity. For example, for a system subject to an external electric field oscillating along one direction with an angular frequency $\omega$, we can obtain the following Green-Kubo formula for the complex electrical conductivity $\sigma(\omega)$ for the system:

$$\sigma(\omega) = \frac{V}{k_B T} \int_0^\infty dt\, e^{-i\omega t} \left\langle \tilde{j}_q ; \tilde{j}_q(t) \right\rangle, \qquad (1.1)$$

where $k_B$ is the Boltzmann constant, $T$ and $V$ are the initial temperature and the volume of the system, and $\left\langle \tilde{j}_q ; \tilde{j}_q(t) \right\rangle$ is the canonical correlation function for the electric current density operator $\tilde{j}_q$ in the interaction picture:

$$\left\langle \tilde{j}_q ; \tilde{j}_q(t) \right\rangle \equiv \frac{1}{\beta} \int_0^\beta du \left\langle \tilde{j}_q(-i\hbar u) \tilde{j}_q(t) \right\rangle_{\text{eq}}, \qquad (1.2)$$

where $\beta \equiv 1/(k_B T)$ and the subscript "eq" indicates that the statistical average is taken when no electric field is applied to the system so that the system is in an equilibrium state at $T$.

For a quantum system in a steady state with a constant current of heat or particles driven by a temperature or chemical potential difference between two reservoirs attached to the system, the fluctuation theorem for the current[1] was previously shown [2, 3, 4] to lead to the Green-Kubo formula for the linear response coefficient for the current expressed in terms of the *symmetrized* correlation functions of the current density operator.

For example, for a quantum system that is subject to a constant magnetic field **B** and is driven to a steady heat conduction state by a temperature difference between two heat reservoirs attached to the system, we can derive the Green-Kubo formula for its thermal conductivity $\kappa$ in terms of the symmetrized correlation function of its heat current density operator $\tilde{j}_{\mathbf{B},\mathrm{F}}^{Q}$ in the Heisenberg picture:

$$\kappa = \frac{V}{k_B \overline{T}^2} \lim_{\tau \to \infty} \frac{1}{\tau} \int_0^{\tau} dt_1 \int_0^{t_1} dt_2 \left\langle \frac{1}{2} \left\{ \tilde{j}_{\mathbf{B},\mathrm{F}}^{Q}(t_1) \tilde{j}_{\mathbf{B},\mathrm{F}}^{Q}(t_2) + \tilde{j}_{\mathbf{B},\mathrm{F}}^{Q}(t_2) \tilde{j}_{\mathbf{B},\mathrm{F}}^{Q}(t_1) \right\} \right\rangle_{\mathrm{eq}} ,$$

$$(1.3)$$

where $V$ is the volume of the system and the subscript "eq" indicates that the statistical average is taken when the temperature difference between the reservoirs is kept at zero so that the system and the reservoirs are all in equilibrium at the same temperature $\overline{T}$.

In this article, we will show that for quantum systems in steady states with constant rates of work done on the systems, we can also derive the fluctuation theorem for a quantity induced in such a system to obtain the Green-Kubo formula expressed in terms of the *symmetrized* correlation function of the induced quantity. The fluctuation theorem for the induced quantity follows from a quantum extension of a general relation (*i.e.*, Equation (1) in [8]) derived by Crooks for classical systems.

---

[1] Earlier, the fluctuation theorem for the current for a classical system had been shown to lead to a Green-Kubo formula expressed in terms of the time correlation function $\left\langle j_q(0) j_q(t) \right\rangle_{\mathrm{eq}}$ of the current density in the system [5, 6, 7].

In Sec. II, as an example of such systems, we will consider a fluid in a steady shear flow driven by a constant velocity of a solid plate moving above the fluid. We will derive the fluctuation theorem for the shear stress on the fluid and obtain the Green-Kubo formula for its shear viscosity $\eta$ in terms of the symmetrized correlation function of its shear stress operator $\tilde{P}_{\mathrm{F}}$ in the Heisenberg picture:

$$\eta = \frac{V}{k_B T} \lim_{\tau \to \infty} \frac{1}{\tau} \int_0^\tau dt_1 \int_0^{t_1} dt_2 \left\langle \frac{1}{2} \left\{ \tilde{P}_{\mathrm{F}}(t_1) \tilde{P}_{\mathrm{F}}(t_2) + \tilde{P}_{\mathrm{F}}(t_2) \tilde{P}_{\mathrm{F}}(t_1) \right\} \right\rangle_{\mathrm{eq}} , \qquad (1.4)$$

where $V$ is the volume of the fluid and the subscript "eq" indicates that the statistical average is taken when the plate remains at rest so that the fluid, the plate, and a heat reservoir attached to the fluid are all in equilibrium at the same temperature $T$.

As steady states in quantum systems are accompanied by either constant rates of work done on the systems or constant currents of heat or particles through the systems, the fluctuation theorem for a quantity induced in such a system therefore always leads to the Green-Kubo formula expressed in terms of the symmetrized correlation function of the induced quantity.

In Appendix, for the sake of completeness, we will derive the fluctuation theorem for the heat current through a quantum system in a steady heat conduction state with a constant heat current driven by a temperature difference between two heat reservoirs attached to the system to obtain the Green-Kubo formula (1.3) for its thermal conductivity $\kappa$. Our derivation of the fluctuation theorem for the heat current is similar to that of the fluctuation theorem for the shear stress in Sec. II.E but different from the existing ones [3, 4, 5] as it employs the quantum extension of the relation by Crooks mentioned above.

# 2. GREEN-KUBO FORMULA FOR SHEAR VISCOSITY

## 2.1. Shear viscosity of a fluid

In this section, as an example of quantum systems in steady states with constant rates of work done on the systems, we consider a fluid driven to a steady shear flow. The fluid is placed between two plates both perpendicular to the $y$-axis and each with a surface area $A$ and the depth of the fluid along the $y$-axis is $h$. The fluid is also attached to a heat reservoir at an inverse temperature $\beta$.

Before an initial time $t = 0$ and after a final time $t = \tau$, the both plates remain at rest. During the time interval $[0, \tau]$, the plate under the fluid remains at rest while the plate above the fluid is moving along the $x$-direction to induce the shear flow in the fluid. We assume that the plate above the fluid moves at a constant velocity $v$ during a time interval $[\tau', \tau - \tau']$, where $\tau'$ is a fixed positive constant satisfying $\tau' << \tau$.

On the macroscopic level, the average work done on the fluid by the moving plate during the time interval $[0, \tau]$ is approximately

$$\overline{W} \cong A\overline{P}_{yx}v\tau, \qquad (2.1)$$

where $\overline{P}_{yx}$ is the average shear stress exerted on the fluid by the plate. For small $|v|$, we define the shear viscosity $\eta$ of the fluid by the following linear response relation:

$$\overline{P}_{yx} \cong \eta \frac{v}{h}. \qquad (2.2)$$

## 2.2. Total Hamiltonian

The total Hamiltonian for the fluid, the plates, and the heat reservoir is given by

$$H(t,v) = H^{(0)} + H^{\mathrm{int}}\left(\{\mathbf{r}_a + X(t)\hat{x}\}_a\right), \qquad (2.3)$$

where $H^{\mathrm{int}}$ is the potential energy due to the interactions between the particles in the fluid and those in the moving plate and $H^{(0)}$ is the rest of the total Hamiltonian. $X(t)$ is the $x$ coordinate of the center of mass of the moving plate and $\mathbf{r}_a$ is the position vector, measured from the center of mass, for the $a$-th atom in the moving plate.

For the fluid and the plates, we impose periodic boundary conditions at the boundaries perpendicular to the $x$ axis. We then assume that $X(t)$ satisfies $X(t) = X(\tau - t)$ and $X(0) = X(\tau) = 0$ so that $H(t,v) = H(\tau - t, v)$ and the moving plate returns to its initial position at the final time $t = \tau$ and

$$H(\tau,v) = H(0,v) = H^{(0)} + H^{\mathrm{int}}\left(\{\mathbf{r}_a\}_a\right) = H(0,0). \qquad (2.4)$$

We also assume that the velocity $\dot{X}(t)$ of the center of mass of the moving plate satisfies $\dot{X}(0) = \dot{X}(\tau) = 0$ and that during the time interval $[\tau', \tau - \tau']$, where $\tau'$ is the fixed constant satisfying $\tau' << \tau$, $\dot{X}(t)$ remains constant at $v$: $\dot{X}(t) = v$.

## 2.3. Initial eigenstates

Before the initial time $t = 0$, we assume that the total system consisting of the fluid, the plates, and the reservoir is in its equilibrium state characterized by the inverse temperature $\beta$. Just before $t = 0$, through a measurement of the energy in the total system, we find the total system to be in an eigenstate $|i\rangle$ of $H(0,0)$ with a corresponding energy eigenvalue $E(i)$:

$$H(0,0)|i\rangle = E(i)|i\rangle. \qquad (2.5)$$

We assume that the initial eigenstate is selected by the following canonical ensemble distribution:

$$\rho_{\mathrm{eq}}(i) \equiv \frac{1}{Z(\beta)} \exp\left[-\beta E(i)\right], \qquad (2.6)$$

where $Z$ is the partition function defined by $Z(\beta) \equiv \sum_i \exp\left[-\beta E(i)\right]$.

## 2.4. Final eigenstates

Just after the final time $t = \tau$, through a measurement of the energy in the total system, we find the total system to be in an eigenstate $|f\rangle$ of $H(0,0)$ with a corresponding energy eigenvalue $E(f)$:

$$H(0,0)|f\rangle = E(f)|f\rangle. \qquad (2.7)$$

Note that the set of all the initial eigenstates is the same as the set of all the final eigenstates: $\{|i\rangle\} = \{|f\rangle\}$.

## 2.5. Time-reversed process and the principle of microreversibility

In this subsection, to make this article to be self-contained, we will derive the principle of microreversibility (2.16) whose direct consequence (2.17) as well as (2.21), (2.22), and (2.23) will be used in Sec. II.E, where we show the fluctuation theorem for the shear stress. The readers who are familiar with these equations may wish to skip this subsection.

### 2.5.1. Principle of microreversibility

During the time interval $[0, \tau]$, the state $\left|\Psi(t)\right\rangle$ of the total system evolves according to the Schrödinger equation with $H(t, v)$ so that its final state is related to its initial state by

$$\left|\Psi(\tau)\right\rangle = U_v \left|\Psi(0)\right\rangle, \tag{2.8}$$

where $U_v$ is the time evolution operator at $t = \tau$. Since the time reversal operator $\Theta$ satisfies $i\Theta = -\Theta i$ and $\Theta\Theta^\dagger = \Theta^\dagger\Theta = I$, the time-reversed state defined by

$$\left|\Psi_r(t)\right\rangle \equiv \Theta\left|\Psi(\tau - t)\right\rangle \tag{2.9}$$

evolves according to the following Schrödinger equation:

$$i\hbar\frac{\partial}{\partial t}\left|\Psi_r(t)\right\rangle = i\hbar\frac{\partial}{\partial t}\left(\Theta\left|\Psi(\tau - t)\right\rangle\right) = \Theta\left\{i\hbar\frac{\partial}{\partial(\tau - t)}\left|\Psi(\tau - t)\right\rangle\right\}$$

$$= \Theta H(\tau - t, v)\left|\Psi(\tau - t)\right\rangle = \Theta H(\tau - t, v)\Theta^\dagger\left(\Theta\left|\Psi(\tau - t)\right\rangle\right)$$

$$= {}^\Theta H(\tau - t, v)\left|\Psi_r(t)\right\rangle,$$

$$(2.10)$$

where the time-reversed Hamiltonian $^{\Theta}H(\tau - t, v)$ is defined by

$$^{\Theta}H(\tau - t, v) \equiv \Theta H(\tau - t, v)\Theta^{\dagger}. \qquad (2.11)$$

We assume our total Hamiltonian satisfies

$$^{\Theta}H(\tau - t, v) = H(t, -v) \qquad (2.12$$

so that $H(0,0)$ is invariant with respect to time-reversal: $^{\Theta}H(0,0) = H(\tau,0) = H(0,0)$.

The final state of the time-reversed backward process is related to its initial state by

$$\left| \Psi_r(\tau) \right\rangle = {}^{\Theta}U_v \left| \Psi_r(0) \right\rangle, \qquad (2.13)$$

where $^{\Theta}U_v$ is the time evolution operator at the end of the backward process, which is controlled by $^{\Theta}H(\tau - t, v) = H(t, -v)$ so that

$$^{\Theta}U_v = U_{-v}. \qquad (2.14)$$

For any $\left| \Psi(0) \right\rangle$, we then find

$$\Theta \left| \Psi(0) \right\rangle = \left| \Psi_r(\tau) \right\rangle = {}^{\Theta}U_v \left| \Psi_r(0) \right\rangle = {}^{\Theta}U_v \Theta \left| \Psi(\tau) \right\rangle = {}^{\Theta}U_v \Theta U_v \left| \Psi(0) \right\rangle \quad (2.15)$$

so that

$$U_{-v} = {}^{\Theta}U_v = \Theta U_v^{\dagger}\Theta^{\dagger}, \qquad (2.16)$$

which is called the principle of microreversibility.

Using this equation, we can show that the transition probability for the forward process from an initial eigenstate $|i\rangle$ to a final eigenstate $|f\rangle$ is equal to the transition probability for the backward time-reversed process from $|^{\Theta}f\rangle \equiv \Theta|f\rangle$ to $|^{\Theta}i\rangle \equiv \Theta|i\rangle$:

$$\left|\langle f|U_\nu|i\rangle\right|^2 = \left|\langle {}^{\Theta}i|{}^{\Theta}U_\nu|{}^{\Theta}f\rangle\right|^2 = \left|\langle {}^{\Theta}i|U_{-\nu}|{}^{\Theta}f\rangle\right|^2, \qquad (2.17)$$

which follows from (2.14) and

$$\left\langle {}^{\Theta}i\big|{}^{\Theta}U_\nu\big|{}^{\Theta}f\right\rangle = \left(\Theta|i\rangle,\ \Theta U_\nu^{\dagger}|f\rangle\right) = \left(U_\nu^{\dagger}|f\rangle,\ |i\rangle\right) = \left\langle f|U_\nu|i\right\rangle, \qquad (2.18)$$

where $\Theta$ is anti-unitary so that for any pair of states, $|\alpha\rangle$ and $|\alpha'\rangle$, $\Theta$ satisfies

$$\left(\Theta|\alpha'\rangle,\ \Theta|\alpha\rangle\right) = \left(|\alpha\rangle,\ |\alpha'\rangle\right), \qquad (2.19)$$

where $\left(|\alpha\rangle,\ |\alpha'\rangle\right)$ is the inner product between $|\alpha\rangle$ and $|\alpha'\rangle$.

### 2.5.2. Useful property of the time reversal operator

The following property of $\Theta$ will be also useful when we show the fluctuation theorem for the shear stress in Sec. II.E. If $|n\rangle$ is an eigenstate of an observable $A$ with a real eigenvalue $a(n)$ so that $A|n\rangle = a(n)|n\rangle$, then $|^{\Theta}n\rangle \equiv \Theta|n\rangle$ is an eigenstate of $^{\Theta}A \equiv \Theta A \Theta^{\dagger}$ with the eigenvalue $a(n)$:

$$^{\Theta}A\left|{}^{\Theta}n\right\rangle = \Theta A \Theta^{\dagger}\Theta|n\rangle = \Theta A|n\rangle = a(n)\Theta|n\rangle = a(n)\left|{}^{\Theta}n\right\rangle. \qquad (2.20)$$

More specifically, $\left|{}^{\Theta}i\right\rangle$ is an eigenstate of $H(0,0){=}{}^{\Theta}H(0,0)$ with the energy eigenvalue $E(i)$:

$$H(0,0)\left|{}^{\Theta}i\right\rangle{=}{}^{\Theta}H(0,0)\left|{}^{\Theta}i\right\rangle = E(i)\left|{}^{\Theta}i\right\rangle \qquad (2.21)$$

and $\left|{}^{\Theta}f\right\rangle$ is an eigenstate of $H(0,0){=}{}^{\Theta}H(0,0)$ with the energy eigenvalue $E(f)$:

$$H(0,0)\left|{}^{\Theta}f\right\rangle{=}{}^{\Theta}H(0,0)\left|{}^{\Theta}f\right\rangle = E(f)\left|{}^{\Theta}f\right\rangle, \qquad (2.22)$$

which also implies

$$\rho_{\mathrm{eq}}\left({}^{\Theta}f\right) = \frac{1}{Z(\beta)}\exp\left[-\beta E(f)\right] = \rho_{\mathrm{eq}}(f), \qquad (2.23)$$

where $\sum_{{}^{\Theta}f}\exp\left[-\beta E(f)\right] = \sum_{f}\exp\left[-\beta E(f)\right] = \sum_{i}\exp\left[-\beta E(i)\right] = Z(\beta)$ because the time reversal operator $\Theta$ provides a one-to-one map from $\left|{}^{\Theta}f\right\rangle$ to $\left|f\right\rangle$ and $\left\{\left|f\right\rangle\right\} = \left\{\left|i\right\rangle\right\}$.

## 2.6. Cumulant generating function for the shear stress

For a forward process from an initial eigenstate $\left|i\right\rangle$ to a final eigenstate $\left|f\right\rangle$, we define the work $W(i,j)$ done on the fluid by

$$W(i,f) \equiv E(f) - E(i) \qquad (2.24)$$

and the corresponding shear stress by

$$P_{yx}(i,f) \equiv \frac{W(i,f)}{Av\tau} = \frac{E(f) - E(i)}{Av\tau}. \tag{2.25}$$

We then define the cumulant generating function for the shear stress by

$$G_P(\lambda_P, v) \equiv -\lim_{\tau \to \infty} \frac{1}{\tau} \ln\left\langle\!\!\left\langle \exp\left[-\tau\lambda_P P_{yx}(i,f)\right]\right\rangle\!\!\right\rangle_v, \tag{2.26}$$

where the transient process average is defined by

$$\left\langle\!\!\left\langle C(i,f)\right\rangle\!\!\right\rangle_v \equiv \sum_{i,f} C(i,f)\left|\langle f|U|i\rangle\right|^2 \rho_{eq}(i). \tag{2.27}$$

In the following, we will assume that the limit in (2.26) exists. Using the cumulant generating function, we can then obtain the average shear stress by

$$\overline{P}_{yx} = \lim_{\tau \to \infty}\left\langle\!\!\left\langle P_{yx}(i,f)\right\rangle\!\!\right\rangle_v = \left.\frac{\partial G_P}{\partial \lambda_P}\right|_{\lambda_P = 0} \tag{2.28}$$

and the shear viscosity by

$$\eta = h\left.\frac{d\overline{P}_{yx}}{dv}\right|_{v=0} = h\left.\frac{\partial^2 G_P}{\partial v \partial \lambda_P}\right|_{\lambda_P = v = 0}. \tag{2.29}$$

$\lim_{\tau \to \infty}\left\langle\!\!\left\langle P_{yx}(i,f)\right\rangle\!\!\right\rangle_v$ can also be written as

$$\lim_{\tau \to \infty}\left\langle\!\!\left\langle P_{yx}(i,f)\right\rangle\!\!\right\rangle_v = \lim_{\tau \to \infty}\mathrm{Tr}\left[\tilde{\rho}_{eq}\left\{\frac{1}{\tau}\int_0^\tau dt\,\tilde{P}_F(t)\right\}\right], \tag{2.30}$$

where we have defined the density matrix $\tilde{\rho}_{\text{eq}}$ corresponding to the initial canonical ensemble distribution $\rho_{\text{eq}}(i)$ (2.6) by

$$\tilde{\rho}_{\text{eq}} \equiv \frac{1}{Z(\beta)}\exp[-\beta H(0,0)]. \qquad (2.31)$$

We have also defined the shear stress operator $\tilde{P}_{\text{F}}$ in the Heisenberg picture by

$$\tilde{P}_{\text{F}}(t) \equiv \frac{1}{A}\sum_a \frac{\partial H_{\text{F}}^{\text{int}}(t,v)}{\partial x_a}, \qquad (2.32)$$

where $H_{\text{F}}^{\text{int}}(t,v) \equiv U_v(t)^{\dagger}H^{\text{int}}(t,v)U_v(t)$ with $U_v(t)$ being the time evolution operator for the total system at $t$ and $x_a$ is the $x$ coordinate of the position vector for the $a$-th atom in the moving plate.

Using $\left|\langle f|U_v|i\rangle\right|^2 = \langle i|U_v^{\dagger}|f\rangle\langle f|U_v|i\rangle$, $\sum_f |f\rangle\langle f| = 1$, and $\tilde{\rho}_{\text{eq}}|i\rangle = \rho_{\text{eq}}(i)|i\rangle$, we can show (2.30) as follows.

$$\lim_{\tau \to \infty} \left\langle\!\!\left\langle \left\langle P_{yx}(i,f) \right\rangle\!\!\right\rangle\right\rangle_v$$

$$= \lim_{\tau \to \infty} \sum_{i,f} \left\{ \frac{E(f) - E(i)}{Av\tau} \right\} \left| \langle f | U_v | i \rangle \right|^2 \rho_{\mathrm{eq}}(i)$$

$$= \lim_{\tau \to \infty} \sum_{i,f} \frac{\langle i | U_v^\dagger H(0,0) | f \rangle \langle f | U_v | i \rangle - \langle i | U_v^\dagger | f \rangle \langle f | U_v H(0,0) | i \rangle}{Av\tau} \rho_{\mathrm{eq}}(i)$$

$$= \lim_{\tau \to \infty} \sum_i \left\langle i \left| \left\{ \frac{U_v^\dagger H(0,0) U_v - H(0,0)}{Av\tau} \right\} \tilde{\rho}_{\mathrm{eq}} \right| i \right\rangle$$

$$= \lim_{\tau \to \infty} \mathrm{Tr}\left[ \tilde{\rho}_{\mathrm{eq}} \left\{ \frac{H_{\mathrm{F}}(\tau,v) - H(0,0)}{Av\tau} \right\} \right]$$

$$= \lim_{\tau \to \infty} \mathrm{Tr}\left[ \tilde{\rho}_{\mathrm{eq}} \frac{1}{\tau} \int_0^\tau dt \frac{1}{A} \sum_a \left( \frac{\partial H_{\mathrm{F}}^{\mathrm{int}}}{\partial x_a} \right) \right] + \lim_{\tau \to \infty} \frac{1}{\tau} \mathrm{Tr}\left[ \tilde{\rho}_{\mathrm{eq}} \int_0^{\tau'} dt \frac{\dot{X}(t) - v}{Av} \sum_a \left( \frac{\partial H_{\mathrm{F}}^{\mathrm{int}}}{\partial x_a} \right) \right]$$

$$\qquad + \lim_{\tau \to \infty} \frac{1}{\tau} \mathrm{Tr}\left[ \tilde{\rho}_{\mathrm{eq}} \int_{\tau - \tau'}^\tau dt \frac{\dot{X}(t) - v}{Av} \sum_a \left( \frac{\partial H_{\mathrm{F}}^{\mathrm{int}}}{\partial x_a} \right) \right]$$

$$= \lim_{\tau \to \infty} \mathrm{Tr}\left[ \tilde{\rho}_{\mathrm{eq}} \left\{ \frac{1}{\tau} \int_0^\tau dt \tilde{P}_{\mathrm{F}}(t) \right\} \right],$$

$$(2.33)$$

where $H_{\mathrm{F}}(t,v) \equiv U_v(t)^\dagger H(t,v) U_v(t)$ and we have also used

$$\frac{H_{\mathrm{F}}(\tau,v) - H(0,0)}{Av\tau}$$

$$= \frac{1}{Av\tau} \int_0^\tau dt \frac{dH_{\mathrm{F}}(t,v)}{dt} = \frac{1}{\tau} \int_0^\tau dt \frac{\dot{X}(t)}{Av} \sum_a \left( \frac{\partial H_{\mathrm{F}}^{\mathrm{int}}}{\partial x_a} \right)$$

$$= \frac{1}{\tau} \int_{\tau'}^{\tau - \tau'} dt \frac{1}{A} \sum_a \left( \frac{\partial H_{\mathrm{F}}^{\mathrm{int}}}{\partial x_a} \right) + \frac{1}{\tau} \int_0^{\tau'} dt \frac{\dot{X}(t)}{Av} \sum_a \left( \frac{\partial H_{\mathrm{F}}^{\mathrm{int}}}{\partial x_a} \right)$$

$$\qquad + \frac{1}{\tau} \int_{\tau - \tau'}^\tau dt \frac{\dot{X}(t)}{Av} \sum_a \left( \frac{\partial H_{\mathrm{F}}^{\mathrm{int}}}{\partial x_a} \right)$$

$$= \frac{1}{\tau} \int_0^\tau dt \frac{1}{A} \sum_a \left( \frac{\partial H_{\mathrm{F}}^{\mathrm{int}}}{\partial x_a} \right) + \frac{1}{\tau} \int_0^{\tau'} dt \frac{\dot{X}(t) - v}{Av} \sum_a \left( \frac{\partial H_{\mathrm{F}}^{\mathrm{int}}}{\partial x_a} \right)$$

$$\qquad + \frac{1}{\tau} \int_{\tau - \tau'}^\tau dt \frac{\dot{X}(t) - v}{Av} \sum_a \left( \frac{\partial H_{\mathrm{F}}^{\mathrm{int}}}{\partial x_a} \right).$$

$$(2.34)$$

## 2.7. Fluctuation theorem for the shear stress

Using (2.17), (2.21), (2.22), and (2.23), we can show the fluctuation theorem for the shear stress,

$$G_P\left(A\beta v - \lambda_P, v\right) = G_P\left(\lambda_P, -v\right), \qquad (2.35)$$

where

$$
\begin{aligned}
G_P\left(\lambda_P, -v\right) &= -\lim_{\tau \to \infty}\frac{1}{\tau}\ln\left\langle\!\!\left\langle\exp\left[-\tau\lambda_P P_{yx}\left({}^\Theta f, {}^\Theta i\right)\right]\right\rangle\!\!\right\rangle_{-v} \\
&= -\lim_{\tau \to \infty}\frac{1}{\tau}\ln\left[\sum_{\Theta_i,\Theta_f}\exp\left[-\tau\lambda_P P_{yx}\left({}^\Theta f, {}^\Theta i\right)\right]\left|\!\left\langle {}^\Theta i\left|U_{-v}\right|{}^\Theta f\right\rangle\!\right|^2\rho_{\mathrm{eq}}\left({}^\Theta f\right)\right].
\end{aligned}
\qquad (2.36)
$$

The fluctuation theorem follows from

$$
\left\langle\!\!\left\langle\exp\left[-\tau\left(A\beta v - \lambda_P\right)P_{yx}(i, f)\right]\right\rangle\!\!\right\rangle_v = \left\langle\!\!\left\langle\exp\left[-\tau\lambda_P P_{yx}\left({}^\Theta f, {}^\Theta i\right)\right]\right\rangle\!\!\right\rangle_{-v},
$$
$$(2.37)$$

which we can show as follows.

$$
\begin{aligned}
&\left\langle\!\!\left\langle\exp\left[-\tau\left(A\beta v - \lambda_P\right)P_{yx}(i, f)\right]\right\rangle\!\!\right\rangle_v \\
&= \sum_{i,f}\exp\left[\tau\lambda_P P_{yx}(i, f)\right]\exp\left[-\beta A v \tau P_{yx}(i, f)\right]\left|\!\left\langle f\left|U_v\right|i\right\rangle\!\right|^2\rho_{\mathrm{eq}}(i) \\
&= \sum_{i,f}\exp\left[-\tau\lambda_P P_{yx}\left({}^\Theta f, {}^\Theta i\right)\right]\frac{\rho_{\mathrm{eq}}\left({}^\Theta f\right)}{\rho_{\mathrm{eq}}(i)}\left|\!\left\langle {}^\Theta i\left|U_{-v}\right|{}^\Theta f\right\rangle\!\right|^2\rho_{\mathrm{eq}}(i) \\
&= \sum_{\Theta_i,\Theta_f}\exp\left[-\tau\lambda_P P_{yx}\left({}^\Theta f, {}^\Theta i\right)\right]\left|\!\left\langle {}^\Theta i\left|U_{-v}\right|{}^\Theta f\right\rangle\!\right|^2\rho_{\mathrm{eq}}\left({}^\Theta f\right) \\
&= \left\langle\!\!\left\langle\exp\left[-\tau\lambda_P P_{yx}\left({}^\Theta f, {}^\Theta i\right)\right]\right\rangle\!\!\right\rangle_{-v},
\end{aligned}
$$

$$(2.38)$$

where we have used (2.17), $\left| \langle f | U_v | i \rangle \right|^2 = \left| \langle ^\Theta i | U_{-v} |^\Theta f \rangle \right|^2$, and

$$P_{yx}(i,f) = \frac{E(f) - E(i)}{Av\tau} = -\frac{E(i) - E(f)}{Av\tau} = -P_{yx}\left(^\Theta f, ^\Theta i\right), \qquad (2.39)$$

where (2.21) and (2.22) are used. We have also used

$$\exp\left[-\beta Av\,\tau P_{yx}(i,f)\right] = \exp\left[-\beta\{E(f) - E(i)\}\right] = \frac{\rho_{\mathrm{eq}}(f)}{\rho_{\mathrm{eq}}(i)} = \frac{\rho_{\mathrm{eq}}\left(^\Theta f\right)}{\rho_{\mathrm{eq}}(i)},$$

$$(2.40)$$

where $\rho_{\mathrm{eq}}\left(^\Theta f\right) = \rho_{\mathrm{eq}}(f)$ (2.23) is used. In (2.38), we have also used the fact that the time reversal operator $\Theta$ provides a one-to-one map from the set of all the eigenstates of $H(0,0)$ to itself and that $\left\{ |^\Theta i \rangle \right\} = \left\{ |i\rangle \right\}$ and $\left\{ |^\Theta f \rangle \right\} = \left\{ |f\rangle \right\}$.

For a general quantum system in a steady state with a constant rate of work, we can generalize (2.37) as

$$\left\langle\!\!\left\langle \exp\left[\tau\lambda\frac{W(i,f)}{\varsigma\tau}\right]\frac{\rho_{\mathrm{eq}}\left(^\Theta f\right)}{\rho_{\mathrm{eq}}(i)} \right\rangle\!\!\right\rangle_\varsigma = \left\langle\!\!\left\langle \exp\left[-\tau\lambda\frac{W(^\Theta f, ^\Theta i)}{\varsigma\tau}\right] \right\rangle\!\!\right\rangle_{\varepsilon_\varsigma\varsigma},$$

$$(2.41)$$

where $\rho_{\mathrm{eq}}\left(^\Theta f\right)\big/\rho_{\mathrm{eq}}(i) = \exp\left[-\beta W(i,f)\right]$ and the work $W(i,f)$ done on the system during a forward process from an initial eigenstate $|i\rangle$ to a final eigenstate $|f\rangle$ is proportional to $\varsigma\tau$, where $\varsigma$ is a quantity, like $Av$, that drives the steady state. We also assume that the total Hamiltonian satisfies $\Theta H(\tau - t, \varsigma)\Theta^\dagger = H\left(t, \varepsilon_\varsigma \varsigma\right)$, where $\varepsilon_\varsigma$ is 1 or $-1$. In Appendix, we will find a similar relation holds for heat current (see (A.40)). We also note that (2.41) is a quantum extension of a general relation (*i.e.*, Equation (1) in [8]) derived by Crooks for classical systems.

## 2.8. Green-Kubo relation

Using the fluctuation theorem for the shear stress (2.35), we find

$$G_P(A\beta v, v) = G_P(0, -v) = 0 \,. \tag{2.42}$$

Using this equation, we can derive the Green-Kubo relation,

$$\eta = \frac{V}{2k_B T} \lim_{\tau \to \infty} \tau \left\langle\!\!\left\langle P_{yx}(i,f)^2 \right\rangle\!\!\right\rangle_v \bigg|_{v=0} \,, \tag{2.43}$$

where $V = Ah$ and $k_B T = 1/\beta$, as follows.

$$
\begin{aligned}
0 &= \frac{\partial^2 G_P(A\beta v, v)}{\partial v^2}\bigg|_{v=0} \\
&= (A\beta)^2 \frac{\partial^2 G_P(\lambda_P, v)}{\partial \lambda_P^2}\bigg|_{\lambda_P = v = 0} + 2A\beta \frac{\partial^2 G_P(\lambda_P, v)}{\partial v \partial \lambda_P}\bigg|_{\lambda_P = v = 0} \\
&\quad + \frac{\partial^2 G_P(\lambda_P, v)}{\partial v^2}\bigg|_{\lambda_P = v = 0} \\
&= -(A\beta)^2 \tau \left\langle\!\!\left\langle P_{yx}(i,f)^2 \right\rangle\!\!\right\rangle_v \bigg|_{v=0} + 2A\beta \frac{\eta}{h} \,,
\end{aligned}
\tag{2.44}
$$

where we have used

$$\frac{\partial^2 G_P(\lambda_P, v)}{\partial v^2}\bigg|_{\lambda_P = v = 0} = \frac{\partial^2 G_P(0, v)}{\partial v^2}\bigg|_{v=0} = 0 \,. \tag{2.45}$$

## 2.9. Green-Kubo formula for the shear viscosity

To derive the Green-Kubo formula for the shear viscosity (1.4) from the Green-Kubo relation (2.43), we first show

$$\lim_{\tau \to \infty} \tau \left\langle\!\left\langle P_{yx}(i,f)^2 \right\rangle\!\right\rangle_v = \lim_{\tau \to \infty} \tau \mathrm{Tr}\left[ \tilde{\rho}_{\mathrm{eq}} \left\{ \frac{1}{\tau} \int_0^\tau dt \tilde{P}_{\mathrm{F}}(t) \right\}^2 \right], \qquad (2.46)$$

Using $\left| \langle f|U_v|i \rangle \right|^2 = \langle i|U_v^\dagger|f \rangle\langle f|U_v|i \rangle$, $\sum_f |f \rangle\langle f| = 1$, $\tilde{\rho}_{\mathrm{eq}}|i \rangle = \rho_{\mathrm{eq}}(i)|i \rangle$, and $\left[ \tilde{\rho}_{\mathrm{eq}}, H(0,0) \right] = 0$, we find

$$\tau \left\langle\!\left\langle P_{yx}(i,f)^2 \right\rangle\!\right\rangle_v$$

$$= \tau \sum_{i,f} \left\{ \frac{E(f)-E(i)}{Av\tau} \right\}^2 \left| \langle f|U_v|i \rangle \right|^2 \rho_{\mathrm{eq}}(i)$$

$$= \tau \sum_{i,f} \left\{ \frac{\langle i|U_v^\dagger H(0,0)^2|f \rangle\langle f|U_v|i \rangle - 2\langle i|U_v^\dagger H(0,0)|f \rangle\langle f|U_v H(0,0)|i \rangle}{(Av\tau)^2} \right.$$

$$\left. + \frac{\langle i|U_v^\dagger|f \rangle\langle f|U_v H(0,0)^2|i \rangle}{(Av\tau)^2} \right\} \rho_{\mathrm{eq}}(i)$$

$$= \tau \sum_i \left\langle i \left| \left\{ \frac{U_v^\dagger H(0,0)^2 U_v - 2U_v^\dagger H(0,0)U_v H(0,0) + H(0,0)^2}{(Av\tau)^2} \right\} \tilde{\rho}_{\mathrm{eq}} \right| i \right\rangle$$

$$= \tau \mathrm{Tr}\left[ \tilde{\rho}_{\mathrm{eq}} \left\{ \frac{H_{\mathrm{F}}(\tau,v) - H(0,0)}{Av\tau} \right\}^2 \right] \xrightarrow[\tau \to \infty]{} \tau \mathrm{Tr}\left[ \tilde{\rho}_{\mathrm{eq}} \left\{ \frac{1}{\tau} \int_0^\tau dt \tilde{P}_{\mathrm{F}}(t) \right\}^2 \right],$$

$$(2.47)$$

where we have also used (2.34).

We then obtain

$$\lim_{\tau \to \infty} \tau \left\langle\!\!\left\langle P_{yx}(i,f)^2 \right\rangle\!\!\right\rangle_v \bigg|_{v=0}$$

$$= \lim_{\tau \to \infty} \tau \mathrm{Tr}\left[ \tilde{\rho}_{\mathrm{eq}} \left\{ \frac{1}{\tau}\int_0^\tau dt \tilde{P}_{\mathrm{F}}(t) \right\}^2 \right]_{v=0}$$

$$= \lim_{\tau \to \infty} \frac{2}{\tau}\int_0^\tau dt_1 \int_0^{t_1} dt_2 \mathrm{Tr}\left[ \tilde{\rho}_{\mathrm{eq}} \frac{1}{2}\left\{ \tilde{P}_{\mathrm{F}}(t_1)\tilde{P}_{\mathrm{F}}(t_2) + \tilde{P}_{\mathrm{F}}(t_2)\tilde{P}_{\mathrm{F}}(t_1) \right\} \right]_{v=0}$$

$$= \lim_{\tau \to \infty} \frac{2}{\tau}\int_0^\tau dt_1 \int_0^{t_1} dt_2 \tilde{C}_P(t_1,t_2) \, .$$

$$(2.48)$$

where we have defined the symmetrized correlation function of the shear stress operator $\tilde{C}_P$ by

$$\tilde{C}_P(t_1,t_2) \equiv \mathrm{Tr}\left[ \tilde{\rho}_{\mathrm{eq}} \frac{1}{2}\left\{ \tilde{P}_{\mathrm{F}}(t_1)\tilde{P}_{\mathrm{F}}(t_2) + \tilde{P}_{\mathrm{F}}(t_2)\tilde{P}_{\mathrm{F}}(t_1) \right\} \right]_{v=0}$$

$$\equiv \left\langle \frac{1}{2}\left\{ \tilde{P}_{\mathrm{F}}(t_1)\tilde{P}_{\mathrm{F}}(t_2) + \tilde{P}_{\mathrm{F}}(t_2)\tilde{P}_{\mathrm{F}}(t_1) \right\} \right\rangle_{\mathrm{eq}} \, .$$

$$(2.49)$$

The subscript "eq" indicates that the statistical average is taken when both of the plates remain at rest so that the fluid, the plates, and the heat reservoir are all in equilibrium at the same temperature $T$.

Using this equation in the Green-Kubo relation (2.43), we finally obtain the Green-Kubo formula for the shear viscosity $\eta$:

$$\eta = \frac{V}{2k_B T} \lim_{\tau \to \infty} \tau \left\langle\!\!\left\langle P_{yx}(i,f)^2 \right\rangle\!\!\right\rangle_v \bigg|_{v=0} = \frac{V}{k_B T} \lim_{\tau \to \infty} \frac{1}{\tau}\int_0^\tau dt_1 \int_0^{t_1} dt_2 \tilde{C}_P(t_1,t_2) \, . .$$

$$(2.50)$$

## 3. CONCLUSIONS

In this article, we have shown that for quantum systems in steady states with constant rates of work done on the systems, the fluctuation theorem for a quantity induced in such a system leads to the Green-Kubo formula for its linear response coefficient expressed in terms of the *symmetrized* correlation function of the induced quantity. As an example of such systems, we have considered a fluid driven to a steady shear flow and derived the fluctuation theorem for the shear stress on the fluid to obtain the Green-Kubo formula for its shear viscosity expressed in terms of the symmetrized correlation function of its shear stress operator.

For a quantum system in a steady state with a constant current of heat or particles driven by a temperature or chemical potential difference between two reservoirs attached to the system, the fluctuation theorem for the current was also previously shown to lead to the Green-Kubo formula for the linear response coefficient for the current expressed in terms of the *symmetrized* correlation functions of the current density operator.

As steady states in quantum systems are accompanied by either constant rates of work done on the systems or constant currents of heat or particles through the systems, the fluctuation theorem for a quantity induced in such a system therefore always leads to the Green-Kubo formula expressed in terms of the symmetrized correlation function of the induced quantity.

## ACKNOWLEDGMENTS

I wish to thank Michele Bock for constant support and encouragement and Richard F. Martin, Jr. and other members of the physics department at Illinois State University for creating a supportive academic environment.

# APPENDIX: GREEN-KUBO FORMULA FOR THERMAL CONDUCTIVITY

## A.1. Thermal conductivity

Consider a system driven to a steady heat conduction state by a temperature difference between two heat reservoirs, A and B. We assume that their temperatures, $T^A$ and $T^B$, satisfy $T^A > T^B$. The inverse temperatures associated with $T^A$ and $T^B$ are defined by $\beta^A \equiv 1/(k_B T^A)$ and $\beta^B \equiv 1/(k_B T^B)$, respectively. We assume that the system and the reservoirs are subject to a constant magnetic field **B**.

We define the thermal conductivity $\kappa$ of the system by the following linear response relation for the average heat current $\bar{J}_Q(\mathbf{B})$:

$$\bar{J}_Q(\mathbf{B}) = \kappa A \frac{T^A - T^B}{L} \equiv \kappa \frac{A k_B \bar{T}^2}{L} \Delta\beta, \qquad (A.1)$$

where $A$ and $L$ are the cross-sectional area and the length of the system. $\Delta\beta$ is defined by

$$\Delta\beta \equiv \beta^B - \beta^A = \frac{1}{k_B T^B} - \frac{1}{k_B T^A} \cong \frac{T^A - T^B}{k_B \bar{T}^2}, \qquad (A.2)$$

where $\bar{T}$ is defined by

$$\frac{1}{k_B \overline{T}} \equiv \overline{\beta} \equiv \frac{\beta^A + \beta^B}{2}, \qquad (A.3)$$

## A.2. Total Hamiltonian

The total Hamiltonian for the system and the reservoirs is given by

$$H_B \equiv H_B^s \otimes I^A \otimes I^B + I^s \otimes H_B^A \otimes I^B + I^s \otimes I^A \otimes H_B^B + H^{int} = H_B^{(0)} + H^{int},$$
$$(A.4)$$

where $H_B^s$ and $H_B^k$ are the Hamiltonians for the system and the $k$-th reservoir ($k$ = A, B) while $H^{int}$ is a *weak* coupling between the system and the reservoirs. Before the initial time $t = 0$ and after the final time $t = \tau$, we set $H^{int} = 0$ so that the system is detached from the reservoirs.

We assume that $H_B$ satisfies

$$H_{-B} = {}^\Theta H_B = \Theta H_B \Theta^\dagger \qquad (A.5)$$

so that

$$U_{-B} = {}^\Theta U_B = \Theta U_B^\dagger \Theta^\dagger. \qquad (A.6)$$

## A.3. Initial eigenstates

Just before $t = 0$, the system is detached from the reservoirs and through a measurement of the energy in the system, we find the system to be in an eigenstate $\left| i_B^s \right\rangle$ of the system Hamiltonian $H_B^s$ with energy eigenvalue $E_B^s\left( i_B^s \right)$. Before $t = 0$, we assume that the system is in an equilibrium state at the inverse temperature $\overline{\beta} = \left(\beta^{\mathrm{A}} + \beta^{\mathrm{B}}\right)\!/2$ so that the initial eigenstate $\left|i_{\mathbf{B}}^{s}\right\rangle$ is selected by the following canonical ensemble distribution:

$$\rho_{\mathbf{B}}^{s}\!\left(i_{\mathbf{B}}^{s}\right) = \frac{1}{Z_{\mathbf{B}}^{s}\!\left(\overline{\beta}\right)} \exp\!\left[-\overline{\beta}E_{\mathbf{B}}^{s}\!\left(i_{\mathbf{B}}^{s}\right)\right], \qquad (\mathrm{A}.7)$$

where $Z_{\mathbf{B}}^{s}$ is the partition function for the initial canonical ensemble for the system. We choose this value $\overline{\beta}$ for the initial inverse temperature for the system so that we can later show the fluctuation theorem for heat current. We assume that after a long time interval, the steady state for the system should become independent of its initial inverse temperature so that we can choose its value almost freely as long as the value is not so different from $\beta^{\mathrm{A}}$ or $\beta^{\mathrm{B}}$.

Just before $t = 0$, through a measurement of the energy in each reservoir, we find the $k$-th reservoir ($k = \mathrm{A,\ B}$) to be in an eigenstate $\left|i_{\mathbf{B}}^{k}\right\rangle$ of the reservoir Hamiltonian $H_{\mathbf{B}}^{k}$ with energy eigenvalue $E_{\mathbf{B}}^{k}\!\left(i_{\mathbf{B}}^{k}\right)$. Before $t = 0$, we assume that the reservoir is in an equilibrium state at inverse temperature $\beta^{k}$ so that the initial eigenstate $\left|i_{\mathbf{B}}^{k}\right\rangle$ is selected by the following canonical ensemble distribution:

$$\rho_{\mathbf{B}}^{k}\!\left(i_{\mathbf{B}}^{k}\right) = \frac{1}{Z_{\mathbf{B}}^{k}\!\left(\beta^{k}\right)} \exp\!\left[-\beta^{k}E_{\mathbf{B}}^{k}\!\left(i_{\mathbf{B}}^{k}\right)\right], \qquad (\mathrm{A}.8)$$

where $Z_{\mathbf{B}}^{k}$ is the partition function for the initial canonical ensemble for the reservoir.

The initial eigenstate $\left|i_{\mathbf{B}}\right\rangle$ for the total system is then $\left|i_{\mathbf{B}}\right\rangle \equiv \left|i_{\mathbf{B}}^{s}\right\rangle \otimes \left|i_{\mathbf{B}}^{\mathrm{A}}\right\rangle \otimes \left|i_{\mathbf{B}}^{\mathrm{B}}\right\rangle$ and the initial canonical ensemble distribution for the total system is

$$\rho_{\mathbf{B}}\!\left(i_{\mathbf{B}}\right) = \rho_{\mathbf{B}}^{s}\!\left(i_{\mathbf{B}}^{s}\right)\!\rho_{\mathbf{B}}^{\mathrm{A}}\!\left(i_{\mathbf{B}}^{\mathrm{A}}\right)\!\rho_{\mathbf{B}}^{\mathrm{B}}\!\left(i_{\mathbf{B}}^{\mathrm{B}}\right). \qquad (\mathrm{A}.9)$$

According to (2.20), $\left|{}^{\Theta}i_{\mathbf{B}}^s\right\rangle \equiv \Theta\left|i_{\mathbf{B}}^s\right\rangle$ is an eigenstate of $H_{-\mathbf{B}}^s = {}^{\Theta}H_{\mathbf{B}}^s$ with the energy eigenvalue $E_{\mathbf{B}}^s\left(i_{\mathbf{B}}^s\right)$ so that

$$H_{-\mathbf{B}}^s\left|{}^{\Theta}i_{\mathbf{B}}^s\right\rangle = E_{\mathbf{B}}^s\left(i_{\mathbf{B}}^s\right)\left|{}^{\Theta}i_{\mathbf{B}}^s\right\rangle, \qquad (A.10)$$

and $\left|{}^{\Theta}i_{\mathbf{B}}^k\right\rangle \equiv \Theta\left|i_{\mathbf{B}}^k\right\rangle$ is an eigenstate of $H_{-\mathbf{B}}^k = {}^{\Theta}H_{\mathbf{B}}^k$ with the energy eigenvalue $E_{\mathbf{B}}^k\left(i_{\mathbf{B}}^k\right)$ so that

$$H_{-\mathbf{B}}^k\left|{}^{\Theta}i_{\mathbf{B}}^k\right\rangle = E_{\mathbf{B}}^k\left(i_{\mathbf{B}}^k\right)\left|{}^{\Theta}i_{\mathbf{B}}^k\right\rangle. \qquad (A.11)$$

## A.4. Final eigenstates

Just after $t = \tau$, the system is detached from the reservoirs and through a measurement of the energy in the system and those in the reservoirs, we find the system to be in an eigenstate $\left|f_{\mathbf{B}}^s\right\rangle$ of the system Hamiltonian $H_{\mathbf{B}}^s$ with energy eigenvalue $E_{\mathbf{B}}^s\left(f_{\mathbf{B}}^s\right)$ while we find the $k$-th reservoir to be in an eigenstate $\left|f_{\mathbf{B}}^k\right\rangle$ of the reservoir Hamiltonian $H_{\mathbf{B}}^k$ with energy eigenvalue $E_{\mathbf{B}}^k\left(f_{\mathbf{B}}^k\right)$. The final eigenstate $\left|f_{\mathbf{B}}\right\rangle$ for the total system is then $\left|f_{\mathbf{B}}\right\rangle \equiv \left|f_{\mathbf{B}}^s\right\rangle \otimes \left|f_{\mathbf{B}}^A\right\rangle \otimes \left|f_{\mathbf{B}}^B\right\rangle$. Note that the set of all the initial eigenstates for the total system is the same as the set of all the final eigenstates: $\left\{\left|i_{\mathbf{B}}\right\rangle\right\} = \left\{\left|f_{\mathbf{B}}\right\rangle\right\}$.

According to (2.20), $\left|{}^{\Theta}f_{\mathbf{B}}^s\right\rangle \equiv \Theta\left|f_{\mathbf{B}}^s\right\rangle$ is an eigenstate of $H_{-\mathbf{B}}^s = {}^{\Theta}H_{\mathbf{B}}^s$ with the energy eigenvalue $E_{\mathbf{B}}^s\left(f_{\mathbf{B}}^s\right)$ so that

$$H_{-\mathbf{B}}^s\left|{}^{\Theta}f_{\mathbf{B}}^s\right\rangle = E_{\mathbf{B}}^s\left(f_{\mathbf{B}}^s\right)\left|{}^{\Theta}f_{\mathbf{B}}^s\right\rangle, \qquad (A.12)$$

and $\left|{}^{\Theta}f_{\mathbf{B}}^k\right\rangle \equiv \Theta\left|f_{\mathbf{B}}^k\right\rangle$ is an eigenstate of $H_{-\mathbf{B}}^k = {}^{\Theta}H_{\mathbf{B}}^k$ with the energy eigenvalue $E_{\mathbf{B}}^k\left(f_{\mathbf{B}}^k\right)$ so that

$$H_{-\mathbf{B}}^{k}\left|{}^{\Theta}f_{\mathbf{B}}^{k}\right\rangle = E_{\mathbf{B}}^{k}\left(f_{\mathbf{B}}^{k}\right)\left|{}^{\Theta}f_{\mathbf{B}}^{k}\right\rangle. \tag{A.13}$$

## A.5. Cumulant generating function for heat current

By applying the first-order time-dependent perturbation theory [9], where we assume $H^{\text{int}}$ to be weak, we find that $\left|\left\langle f_{\mathbf{B}}\left|U_{\mathbf{B}}\right|i_{\mathbf{B}}\right\rangle\right|^{2}$ is appreciable only when

$$\left|\left\{E_{\mathbf{B}}^{s}\left(f_{\mathbf{B}}^{s}\right) + E_{\mathbf{B}}^{A}\left(f_{\mathbf{B}}^{A}\right) + E_{\mathbf{B}}^{B}\left(f_{\mathbf{B}}^{B}\right)\right\} - \left\{E_{\mathbf{B}}^{s}\left(i_{\mathbf{B}}^{s}\right) + E_{\mathbf{B}}^{A}\left(i_{\mathbf{B}}^{A}\right) + E_{\mathbf{B}}^{B}\left(i_{\mathbf{B}}^{B}\right)\right\}\right| < \frac{2\pi}{\tau}. \tag{A.14}$$

For sufficiently long $\tau$, we can then assume that

$$\Delta E_{\mathbf{B}}^{s} + \Delta E_{\mathbf{B}}^{A} + \Delta E_{\mathbf{B}}^{B} = 0 , \tag{A.15}$$

where $\Delta E_{\mathbf{B}}^{s}$ for the system is defined by

$$\Delta E_{\mathbf{B}}^{s} \equiv E_{\mathbf{B}}^{s}\left(f_{\mathbf{B}}^{s}\right) - E_{\mathbf{B}}^{s}\left(i_{\mathbf{B}}^{s}\right) \tag{A,16}$$

and $\Delta E_{\mathbf{B}}^{k}$ for the $k$-th reservoir is defined by

$$\Delta E_{\mathbf{B}}^{k} \equiv E_{\mathbf{B}}^{k}\left(f_{\mathbf{B}}^{k}\right) - E_{\mathbf{B}}^{k}\left(i_{\mathbf{B}}^{k}\right). \tag{A.17}$$

For a forward process for the total system from an initial eigenstate $\left|i_{\mathbf{B}}\right\rangle$ to a final eigenstate $\left|f_{\mathbf{B}}\right\rangle$, we define the heat transferred from the $k$-th reservoir into the system by

$$Q_{\mathbf{B}}^k\left(i_{\mathbf{B}}^k, f_{\mathbf{B}}^k\right) \equiv -\Delta E_{\mathbf{B}}^k \qquad (A.18)$$

and the average heat current through the system by

$$J_{\mathbf{B}}^Q\left(i_{\mathbf{B}}, f_{\mathbf{B}}\right) \equiv \frac{1}{2}\left[\frac{Q_{\mathbf{B}}^{\mathrm{A}}\left(i_{\mathbf{B}}^{\mathrm{A}}, f_{\mathbf{B}}^{\mathrm{A}}\right)}{\tau} + \frac{\left\{-Q_{\mathbf{B}}^{\mathrm{B}}\left(i_{\mathbf{B}}^{\mathrm{B}}, f_{\mathbf{B}}^{\mathrm{B}}\right)\right\}}{\tau}\right]$$

$$= \frac{-\left\{E_{\mathbf{B}}^{\mathrm{A}}\left(f_{\mathbf{B}}^{\mathrm{A}}\right) - E_{\mathbf{B}}^{\mathrm{A}}\left(i_{\mathbf{B}}^{\mathrm{A}}\right)\right\} + \left\{E_{\mathbf{B}}^{\mathrm{B}}\left(f_{\mathbf{B}}^{\mathrm{B}}\right) - E_{\mathbf{B}}^{\mathrm{B}}\left(i_{\mathbf{B}}^{\mathrm{B}}\right)\right\}}{2\tau} \;.$$

$$(A.19)$$

We then define the cumulant generating function for the heat current by

$$G_Q\left(\lambda_Q, \Delta\beta; \mathbf{B}\right) \equiv -\lim_{\tau \to \infty}\frac{1}{\tau}\ln\left\langle\!\left\langle\exp\left[-\tau\lambda_Q J_{\mathbf{B}}^Q\left(i_{\mathbf{B}}, f_{\mathbf{B}}\right)\right]\right\rangle\!\right\rangle_{\mathbf{B}}, \qquad (A.20)$$

where the transient process average is defined by

$$\left\langle\!\left\langle C\left(i_{\mathbf{B}}, f_{\mathbf{B}}\right)\right\rangle\!\right\rangle_{\mathbf{B}} \equiv \sum_{i_{\mathbf{B}}, f_{\mathbf{B}}} C\left(i_{\mathbf{B}}, f_{\mathbf{B}}\right)\left|\left\langle f_{\mathbf{B}}\left|U_{\mathbf{B}}\right|i_{\mathbf{B}}\right\rangle\right|^2 \rho_{\mathbf{B}}\left(i_{\mathbf{B}}\right). \qquad (A.21)$$

In the following, we will assume that the limit in (A.20) exists. Using the cumulant generating function, we can then obtain the average heat current by

$$\bar{J}_Q(\mathbf{B}) = \lim_{\tau \to \infty}\left\langle\!\left\langle J_{\mathbf{B}}^Q\right\rangle\!\right\rangle_{\mathbf{B}} = \left.\frac{\partial G_Q}{\partial \lambda_Q}\right|_{\lambda_Q = 0} \qquad (A.22)$$

and the thermal conductivity by

$$\kappa = \left(\frac{L}{Ak_B\overline{T}^2}\right)\left.\frac{\partial \bar{J}_Q(\mathbf{B})}{\partial \Delta\beta}\right|_{\Delta\beta = 0} = \left(\frac{L}{Ak_B\overline{T}^2}\right)\left.\frac{\partial^2 G_Q}{\partial \Delta\beta \partial \lambda_Q}\right|_{\lambda_Q = \Delta\beta = 0} \;. \qquad (A.23)$$

$\left\langle\!\left\langle J_{\mathbf{B}}^{\mathcal{Q}}\right\rangle\!\right\rangle_{\mathbf{B}}$ can also be written as

$$\lim_{\tau\to\infty}\left\langle\!\left\langle J_{\mathbf{B}}^{\mathcal{Q}}\right\rangle\!\right\rangle_{\mathbf{B}}=\lim_{\tau\to\infty}\mathrm{Tr}\left[\tilde{\rho}_{\mathbf{B}}\left\{\frac{1}{\tau}\int_0^\tau dt\,\tilde{J}_{\mathbf{B},\mathrm{F}}^{\mathcal{Q}}(t)\right\}\right], \qquad (A.24)$$

where $\tilde{\rho}_{\mathbf{B}}$, the density matrix corresponding to the initial canonical ensemble distribution for the total system (A.9) is defined by

$$\tilde{\rho}_{\mathbf{B}}\equiv\frac{1}{Z_{\mathbf{B}}^{\mathrm{s}}\left(\overline{\beta}\right)}\exp\left(-\overline{\beta}H_{\mathbf{B}}^{\mathrm{s}}\right)\otimes\frac{1}{Z_{\mathbf{B}}^{\mathrm{A}}\left(\beta^{\mathrm{A}}\right)}\exp\left(-\beta^{\mathrm{A}}H_{\mathbf{B}}^{\mathrm{A}}\right)\otimes\frac{1}{Z_{\mathbf{B}}^{\mathrm{B}}\left(\beta^{\mathrm{B}}\right)}\exp\left(-\beta^{\mathrm{B}}H_{\mathbf{B}}^{\mathrm{B}}\right).$$

$$(A.25)$$

We have also defined the heat current operator $\tilde{J}_{\mathbf{B},\mathrm{F}}^{\mathcal{Q}}$ in the Heisenberg picture by

$$\tilde{J}_{\mathbf{B},\mathrm{F}}^{\mathcal{Q}}(t)\equiv\frac{1}{2}\left(-\frac{dH_{\mathbf{B},\mathrm{F}}^{\mathrm{A}}}{dt}+\frac{dH_{\mathbf{B},\mathrm{F}}^{\mathrm{B}}}{dt}\right), \qquad (A.26)$$

where $H_{\mathbf{B},\mathrm{F}}^{k}(t)\equiv U_{\mathbf{B}}(t)^{\dagger}H_{\mathbf{B}}^{k}U_{\mathbf{B}}(t)$ with $U_{\mathbf{B}}(t)$ being the time evolution operator for the total system at $t$ and

$$i\hbar\frac{dH_{\mathbf{B},\mathrm{F}}^{k}}{dt}=\left[H_{\mathbf{B},\mathrm{F}}^{k}(t),H_{\mathbf{B},\mathrm{F}}(t)\right] \qquad (A.27)$$

so that

$$\int_0^\tau dt\,\tilde{J}_{\mathbf{B},\mathrm{F}}^{\mathcal{Q}}(t)=\frac{1}{2}\left[-\left\{H_{\mathbf{B},\mathrm{F}}^{\mathrm{A}}(\tau)-H_{\mathbf{B}}^{\mathrm{A}}\right\}+\left\{H_{\mathbf{B},\mathrm{F}}^{\mathrm{B}}(\tau)-H_{\mathbf{B}}^{\mathrm{B}}\right\}\right]. \qquad (A.28)$$

Recalling $Q_{\mathbf{B}}^k = -\left\{E_{\mathbf{B}}^k\left(f_{\mathbf{B}}^k\right) - E_{\mathbf{B}}^k\left(i_{\mathbf{B}}^k\right)\right\}$ (A.18) and using $\left|\left\langle f_{\mathbf{B}}\left|U_{\mathbf{B}}\right|i_{\mathbf{B}}\right\rangle\right|^2 = \left\langle i_{\mathbf{B}}\left|U_{\mathbf{B}}^\dagger\right|f_{\mathbf{B}}\right\rangle\left\langle f_{\mathbf{B}}\left|U_{\mathbf{B}}\right|i_{\mathbf{B}}\right\rangle$, $\sum_{f_{\mathbf{B}}}\left|f_{\mathbf{B}}\right\rangle\left\langle f_{\mathbf{B}}\right| = 1$, and $\tilde{\rho}_{\mathbf{B}}\left|i_{\mathbf{B}}\right\rangle = \rho_{\mathbf{B}}(i_{\mathbf{B}})\left|i_{\mathbf{B}}\right\rangle$, we can show (A.24):

$$
\begin{aligned}
&\left\langle\!\left\langle J_{\mathbf{B}}^Q(i_{\mathbf{B}}, f_{\mathbf{B}})\right\rangle\!\right\rangle_{\mathbf{B}} \\
&= \sum_{i,f}\left\{\frac{-\left\{E_{\mathbf{B}}^{\mathrm{A}}\left(f_{\mathbf{B}}^{\mathrm{A}}\right) - E_{\mathbf{B}}^{\mathrm{A}}\left(i_{\mathbf{B}}^{\mathrm{A}}\right)\right\} + \left\{E_{\mathbf{B}}^{\mathrm{B}}\left(f_{\mathbf{B}}^{\mathrm{B}}\right) - E_{\mathbf{B}}^{\mathrm{B}}\left(i_{\mathbf{B}}^{\mathrm{B}}\right)\right\}}{2\tau}\right\}\left|\left\langle f_{\mathbf{B}}\left|U_{\mathbf{B}}\right|i_{\mathbf{B}}\right\rangle\right|^2 \rho_{\mathbf{B}}(i_{\mathbf{B}}) \\
&= \mathrm{Tr}\left[\tilde{\rho}_{\mathbf{B}}\left[\frac{-\left\{H_{\mathbf{B},\mathrm{F}}^{\mathrm{A}}(\tau) - H_{\mathbf{B}}^{\mathrm{A}}\right\} + \left\{H_{\mathbf{B},\mathrm{F}}^{\mathrm{B}}(\tau) - H_{\mathbf{B}}^{\mathrm{B}}\right\}}{2\tau}\right]\right] \\
&= \mathrm{Tr}\left[\tilde{\rho}_{\mathbf{B}}\left\{\frac{1}{\tau}\int_0^\tau dt\, \tilde{J}_{\mathbf{B},\mathrm{F}}^Q(t)\right\}\right].
\end{aligned}
$$

(A.29)

## A.6. Fluctuation theorem for the heat current

We can show the fluctuation theorem for the heat current,

$$
G_Q\left(\Delta\beta - \lambda_Q, \Delta\beta; \mathbf{B}\right) = G_Q\left(\lambda_Q, \Delta\beta; -\mathbf{B}\right),
$$

(A.30)

where

$$
\begin{aligned}
&G_Q\left(\lambda_Q, \Delta\beta; -\mathbf{B}\right) \\
&= -\lim_{\tau\to\infty}\frac{1}{\tau}\ln\left\langle\!\left\langle\exp\left[-\tau\lambda_Q J_{-\mathbf{B}}^Q\left(^\Theta f_{\mathbf{B}}, {}^\Theta i_{\mathbf{B}}\right)\right]\right\rangle\!\right\rangle_{-\mathbf{B}}. \\
&= -\lim_{\tau\to\infty}\frac{1}{\tau}\ln\left[\sum_{\Theta i,\Theta f}\exp\left[\tau\lambda_Q J_{-\mathbf{B}}^Q\left(^\Theta f_{\mathbf{B}}, {}^\Theta i_{\mathbf{B}}\right)\right]\left|\left\langle^\Theta i_{\mathbf{B}}\left|U_{-\mathbf{B}}\right|^\Theta f_{\mathbf{B}}\right\rangle\right|^2 \rho_{-\mathbf{B}}\left(^\Theta f_{\mathbf{B}}\right)\right].
\end{aligned}
$$

(A.31)

The fluctuation theorem follows from

$$\left\langle\!\!\left\langle \exp\!\left[-\tau\!\left(\Delta\beta-\lambda_Q\right)J_\mathbf{B}^Q\!\left(i_\mathbf{B},f_\mathbf{B}\right)\right]\right\rangle\!\!\right\rangle_\mathbf{B} = \left\langle\!\!\left\langle \exp\!\left[-\tau\lambda_Q J_{-\mathbf{B}}^Q\!\left(^\Theta f_\mathbf{B},{}^\Theta i_\mathbf{B}\right)\right]\right\rangle\!\!\right\rangle_{-\mathbf{B}},$$

$$(A.32)$$

which we can show as follows.

$$\left\langle\!\!\left\langle \exp\!\left[-\tau\!\left(\Delta\beta-\lambda_Q\right)J_\mathbf{B}^Q\!\left(i_\mathbf{B},f_\mathbf{B}\right)\right]\right\rangle\!\!\right\rangle_\mathbf{B}$$

$$= \sum_{i,f} \exp\!\left[\tau\lambda_Q J_\mathbf{B}^Q\!\left(i_\mathbf{B},f_\mathbf{B}\right)\right]\exp\!\left[-\Delta\beta\tau J_\mathbf{B}^Q\!\left(i_\mathbf{B},f_\mathbf{B}\right)\right]\!\left|\!\left\langle f_\mathbf{B}\left|U_\mathbf{B}\right|i_\mathbf{B}\right\rangle\!\right|^2 \rho_\mathbf{B}\!\left(i_\mathbf{B}\right)$$

$$= \sum_{i,f} \exp\!\left[\tau\lambda_Q J_{-\mathbf{B}}^Q\!\left(^\Theta f_\mathbf{B},{}^\Theta i_\mathbf{B}\right)\right]\frac{\rho_{-\mathbf{B}}\!\left(^\Theta f_\mathbf{B}\right)}{\rho_\mathbf{B}\!\left(i_\mathbf{B}\right)}\!\left|\!\left\langle ^\Theta i_\mathbf{B}\left|U_{-\mathbf{B}}\right|^\Theta f_\mathbf{B}\right\rangle\!\right|^2 \rho_\mathbf{B}\!\left(i_\mathbf{B}\right)$$

$$= \sum_{^\Theta i,\,^\Theta f} \exp\!\left[\tau\lambda_Q J_{-\mathbf{B}}^Q\!\left(^\Theta f_\mathbf{B},{}^\Theta i_\mathbf{B}\right)\right]\!\left|\!\left\langle ^\Theta i_\mathbf{B}\left|U_{-\mathbf{B}}\right|^\Theta f_\mathbf{B}\right\rangle\!\right|^2 \rho_{-\mathbf{B}}\!\left(^\Theta f_\mathbf{B}\right)$$

$$= \left\langle\!\!\left\langle \exp\!\left[-\tau\lambda_Q J_{-\mathbf{B}}^Q\!\left(^\Theta f_\mathbf{B},{}^\Theta i_\mathbf{B}\right)\right]\right\rangle\!\!\right\rangle_{-\mathbf{B}}.$$

$$(A.33)$$

In this derivation, we have used a consequence of the principle of microreversibility,

$$\left|\!\left\langle f_\mathbf{B}\left|U_\mathbf{B}\right|i_\mathbf{B}\right\rangle\!\right|^2 = \left|\!\left\langle ^\Theta i_\mathbf{B}\left|^\Theta U_\mathbf{B}\right|^\Theta f_\mathbf{B}\right\rangle\!\right|^2 = \left|\!\left\langle ^\Theta i_\mathbf{B}\left|U_{-\mathbf{B}}\right|^\Theta f_\mathbf{B}\right\rangle\!\right|^2, \qquad (A.34)$$

and

$$J_\mathbf{B}^Q\!\left(i_\mathbf{B},f_\mathbf{B}\right) = \frac{-\left\{E_\mathbf{B}^\mathrm{A}\!\left(f_\mathbf{B}^\mathrm{A}\right)-E_\mathbf{B}^\mathrm{A}\!\left(i_\mathbf{B}^\mathrm{A}\right)\right\}+\left\{E_\mathbf{B}^\mathrm{B}\!\left(f_\mathbf{B}^\mathrm{B}\right)-E_\mathbf{B}^\mathrm{B}\!\left(i_\mathbf{B}^\mathrm{B}\right)\right\}}{2\,\tau} = -J_{-\mathbf{B}}^Q\!\left(^\Theta f_\mathbf{B},{}^\Theta i_\mathbf{B}\right),$$

$$(A.35)$$

where (A.11) and (A.13) are used. We have also used

$$\exp\left[-\tau\Delta\beta J_{\mathbf{B}}^{Q}(i_{\mathbf{B}}, f_{\mathbf{B}})\right] = \exp\left(-\Delta\beta\frac{-\Delta E_{\mathbf{B}}^{A} + \Delta E_{\mathbf{B}}^{B}}{2}\right)$$

$$= \exp\left[-\overline{\beta}\left(\Delta E_{\mathbf{B}}^{s} + \Delta E_{\mathbf{B}}^{A} + \Delta E_{\mathbf{B}}^{B}\right) - \Delta\beta\frac{-\Delta E_{\mathbf{B}}^{A} + \Delta E_{\mathbf{B}}^{B}}{2}\right]$$

$$= \exp\left(-\overline{\beta}\Delta E_{\mathbf{B}}^{s} - \beta^{A}\Delta E_{\mathbf{B}}^{A} - \beta^{B}\Delta E_{\mathbf{B}}^{B}\right)$$

$$= \frac{\rho_{\mathbf{B}}(f_{\mathbf{B}})}{\rho_{\mathbf{B}}(i_{\mathbf{B}})} = \frac{\rho_{-\mathbf{B}}(^{\Theta}f_{\mathbf{B}})}{\rho_{\mathbf{B}}(i_{\mathbf{B}})} ,$$

$$(A.36)$$

where we have used (A.15), $\Delta E_{\mathbf{B}}^{s} + \Delta E_{\mathbf{B}}^{A} + \Delta E_{\mathbf{B}}^{B} = 0$, as well as

$$\beta^{A} = \frac{\beta^{A} + \beta^{B}}{2} - \frac{\beta^{B} - \beta^{A}}{2} = \overline{\beta} - \frac{1}{2}\Delta\beta \qquad (A.37)$$

and

$$\beta^{B} = \frac{\beta^{A} + \beta^{B}}{2} + \frac{\beta^{B} - \beta^{A}}{2} = \overline{\beta} + \frac{1}{2}\Delta\beta. \qquad (A.38)$$

We have also used

$$\rho_{-\mathbf{B}}\left(^{\Theta}f_{\mathbf{B}}\right) = \rho_{\mathbf{B}}(f_{\mathbf{B}}), \qquad (A.39)$$

which follows from (A.12) and (A.13).

(A.32) can be written as

$$\left\langle\!\left\langle\exp\left[\tau\lambda_{Q}J_{\mathbf{B}}^{Q}(i_{\mathbf{B}}, f_{\mathbf{B}})\right]\frac{\rho_{-\mathbf{B}}\left(^{\Theta}f_{\mathbf{B}}\right)}{\rho_{\mathbf{B}}(i_{\mathbf{B}})}\right\rangle\!\right\rangle_{\mathbf{B}} = \left\langle\!\left\langle\exp\left[-\tau\lambda_{Q}J_{-\mathbf{B}}^{Q}\left(^{\Theta}f_{\mathbf{B}}, {}^{\Theta}i_{\mathbf{B}}\right)\right]\right\rangle\!\right\rangle_{-\mathbf{B}},$$

$$(A.40)$$

which is similar to (2.41) for the shear stress.

## A.7. Green-Kubo formula for the thermal conductivity

Using the fluctuation theorem for the heat current (A.30), we find

$$G_Q(\Delta\beta, \Delta\beta; \mathbf{B}) = G(0, \Delta\beta; -\mathbf{B}) = 0 \,.\qquad\text{(A.41)}$$

Using this equation, we can derive the Green-Kubo relation,

$$\kappa = \frac{L}{2Ak_B\bar{T}^2}\lim_{\tau\to\infty}\tau\left\langle\!\left\langle J_{\mathbf{B}}^Q(i_{\mathbf{B}}, f_{\mathbf{B}})^2\right\rangle\!\right\rangle_{\mathbf{B},\Delta\beta=0},\qquad\text{(A.42)}$$

which follows from

$$
\begin{aligned}
0 &= \left.\frac{\partial^2 G_Q(\Delta\beta, \Delta\beta; \mathbf{B})}{\partial\Delta\beta^2}\right|_{\Delta\beta=0}\\
&= \left.\frac{\partial^2 G_Q(\lambda_Q, \Delta\beta; \mathbf{B})}{\partial\lambda_Q^2}\right|_{\lambda_Q=\Delta\beta=0} + 2\left.\frac{\partial^2 G_Q(\lambda_Q, \Delta\beta; \mathbf{B})}{\partial\Delta\beta\partial\lambda_Q}\right|_{\lambda_Q=\Delta\beta=0}\\
&\quad + \left.\frac{\partial^2 G_Q(\lambda_Q, \Delta\beta; \mathbf{B})}{\partial\Delta\beta^2}\right|_{\lambda_Q=\Delta\beta=0}\\
&= -\tau\left\langle\!\left\langle J_{\mathbf{B}}^Q(i_{\mathbf{B}}, f_{\mathbf{B}})^2\right\rangle\!\right\rangle_{\mathbf{B},\Delta\beta=0} + \frac{2Ak_B T^2}{L}\kappa,
\end{aligned}
$$

$$\text{(A.43)}$$

where we have used (A.20) and (A.23).

Recalling $Q_{\mathbf{B}}^k = -\left\{E_{\mathbf{B}}^k(f_{\mathbf{B}}^k) - E_{\mathbf{B}}^k(i_{\mathbf{B}}^k)\right\}$ (A.18) and using $\left[\tilde{\rho}_{\mathbf{B}}, H_{\mathbf{B}}^k\right] = 0$ and $\left[H_{\mathbf{B}}^{\mathrm{A}}, H_{\mathbf{B}}^{\mathrm{B}}\right] = \left[H_{\mathbf{B},\mathrm{F}}^{\mathrm{A}}(\tau), H_{\mathbf{B},\mathrm{F}}^{\mathrm{B}}(\tau)\right] = 0$, we can then show

$$\left\langle\!\!\left\langle J_{\mathbf{B}}^{Q}(i_{\mathbf{B}}, f_{\mathbf{B}})^2 \right\rangle\!\!\right\rangle_{\mathbf{B},\Delta\beta=0} = \mathrm{Tr}\left[\tilde{\rho}_{\mathbf{B}}\left[\frac{-\left\{H_{\mathbf{B},\mathrm{F}}^{\mathrm{A}}(\tau)-H_{\mathbf{B}}^{\mathrm{A}}\right\}+\left\{H_{\mathbf{B},\mathrm{F}}^{\mathrm{B}}(\tau)-H_{\mathbf{B}}^{\mathrm{B}}\right\}}{2\tau}\right]^2\right]_{\Delta\beta=0}$$

$$= \mathrm{Tr}\left[\tilde{\rho}_{\mathbf{B}}\left\{\frac{1}{\tau}\int_0^\tau dt\,\tilde{J}_{\mathbf{B},\mathrm{F}}^{Q}(t)\right\}^2\right]_{\Delta\beta=0} .$$

$$(A.44)$$

Using this equation in the Green-Kubo relation (A.42), we obtain

$$\kappa = \frac{L}{2Ak_B\overline{T}^2}\lim_{\tau\to\infty}\tau\left\langle\!\!\left\langle J_{\mathbf{B}}^{Q}(i_{\mathbf{B}}, f_{\mathbf{B}})^2 \right\rangle\!\!\right\rangle_{\mathbf{B},\Delta\beta=0}$$

$$= \frac{L}{2Ak_B\overline{T}^2}\lim_{\tau\to\infty}\tau\mathrm{Tr}\left[\tilde{\rho}_{\mathbf{B}}\left\{\frac{1}{\tau}\int_0^\tau dt\,\tilde{J}_{\mathbf{B},\mathrm{F}}^{Q}(t)\right\}^2\right]_{\Delta\beta=0}$$

$$= \frac{V}{2k_B\overline{T}^2}\lim_{\tau\to\infty}\frac{1}{\tau}\mathrm{Tr}\left[\tilde{\rho}_{\mathbf{B}}\left\{\int_0^\tau dt\,\tilde{j}_{\mathbf{B},\mathrm{F}}^{Q}(t)\right\}^2\right]_{\Delta\beta=0} ,$$

$$(A.45)$$

where $V = AL$ and we have defined the heat current density operator $\tilde{j}_{\mathbf{B},\mathrm{F}}^{Q}$ in the Heisenberg picture by

$$\tilde{j}_{\mathbf{B},\mathrm{F}}^{Q}(t) \equiv \frac{1}{A}\tilde{J}_{\mathbf{B},\mathrm{F}}^{Q}(t). \qquad (A.46)$$

Following steps similar to those for the shear viscosity, we then obtain the Green-Kubo formula for $\kappa$,

$$\kappa = \frac{V}{k_B\overline{T}^2}\lim_{\tau\to\infty}\frac{1}{\tau}\int_0^\tau dt_1\int_0^{t_1} dt_2\,\tilde{C}_{\mathbf{B}}^{Q}(t_1, t_2), \qquad (A.47)$$

where the symmetrized correlation function of the heat current density operator is defined by

$$\tilde{C}_{\mathbf{B}}^{Q}(t_1,t_2) \equiv \mathrm{Tr}\left[\tilde{\rho}_{\mathbf{B}}\frac{1}{2}\left\{\tilde{j}_{\mathbf{B},\mathrm{F}}^{Q}(t_1)\tilde{j}_{\mathbf{B},\mathrm{F}}^{Q}(t_2) + \tilde{j}_{\mathbf{B},\mathrm{F}}^{Q}(t_2)\tilde{j}_{\mathbf{B},\mathrm{F}}^{Q}(t_1)\right\}\right]_{\Delta\beta=0}$$

$$\equiv \left\langle\frac{1}{2}\left\{\tilde{j}_{\mathbf{B},\mathrm{F}}^{Q}(t_1)\tilde{j}_{\mathbf{B},\mathrm{F}}^{Q}(t_2) + \tilde{j}_{\mathbf{B},\mathrm{F}}^{Q}(t_2)\tilde{j}_{\mathbf{B},\mathrm{F}}^{Q}(t_1)\right\}\right\rangle_{\mathrm{eq}} .$$

$$(A.48)$$

The subscript "eq" indicates that the statistical average is taken when the temperature difference between the reservoirs is kept at zero so that $\Delta\beta = 0$ and the system and the reservoirs are all in equilibrium at the same temperature $\bar{T}$.

# REFERENCES

1,   R. Kubo, M. Toda, and N. Hashitsume, *Statistical Physics II: Nonequilibrium Statistical Mechanics* (Springer-Verlag, Berlin, 1991), Sec. 4.2.2.

2,   K. Saito and A. Dhar, *Fluctuation Theorem in Quantum Heat Conduction,* Phys. Rev. Lett. **99**, 180601(2007).

3.   K. Saito and Y. Utsumi, *Symmetry in full counting statistics, fluctuation theorem, and relations among nonlinear transport coefficients in the presence of a magnetic field,* Phys. Rev. B **78**,115429(2008).

4.   D. Andrieux, P. Gaspard, T. Monnai, and S. Tasaki, *Fluctuation theorem for currents in open quantum systems*, New J. Phys. **11**, 043014 (2009).

5.   G. Gallavotti, *Extension of Onsager's reciprocity to large fields and chaotic hypothesis*, Phys. Rev. Lett. **77**, 4334 (1996).

6.   J. L. Lebowitz and H. Spohn, *A Gallavotti-Cohen-Type Symmetry in the Large Deviation Functional for Stochastic Dynamics*, J. Stat. Phys. **95**, 333 (1999).

7.   C. Maes, *The Fluctuation Theorem as a Gibbs Property,* J. Stat. Phys. **95**, 367 (1999).

8.   G. E. Crooks, *Path-ensemble averages in systems driven far from equilibrium*, Phys. Rev. E **61**, 2361 (2000).

9.   A. Messiah, *Quantum Mechanics* (North-Holland, Amsterdam, 1962), Ch. XVII, Sec. 4.